# High–Speed Imagery Analysis Of Droplet Impact On Soft Oil-Infused Surface.

*Shubham S. Ganar[1], Deepak J.[1] and Arindam Das[1]\**

[1]*School of Mechanical Sciences, Indian Institute of Technology (IIT) Goa, GEC Campus, Farmagudi, Ponda, Goa 403401, India*

## ABSTRACT

Droplet impact on solid liquid-infused surfaces (LIS) has been widely explored due to its significant scientific implications and industrial relevance. In most studies, the predominant impact behavior observed is complete droplet rebound. In This study we investigated the influence of octadecyltrichlorosilane (OTS) functionalization and oil coatings on the droplet impact dynamics of smooth polydimethylsiloxane (PDMS) surfaces. We conducted droplet impact experiments on smooth PDMS functionalized with OTS and subsequently coated or absorbed with two different oils—silicone oil (5cSt) and hexadecane—to create Van der Waals and non-Van der Waals SLIP surfaces. Contact angle measurements revealed that OTS functionalization reduced adhesion and increased water repellency, facilitating partial droplet rebound upon impact. Oil-coated surfaces exhibited reduced droplet spreading due to viscous resistance, while absorbed oils altered surface flexibility, influencing impact dynamics. PDMS samples absorbed with silicone oil demonstrated complete droplet rebound at all Weber numbers, whereas hexadecane-absorbed surfaces exhibited limited spreading and no rebound, highlighting the significance of oil-PDMS interactions. High-speed imaging and quantitative analysis confirmed that surface functionalization and oil interactions critically affect droplet spreading, recoil, and rebound behavior. These findings provide insights into optimizing liquid-repellent surfaces for applications such as self-cleaning coatings and droplet transport systems.

**Corresponding Author**

Arindam Das*, Associate Professor, School of Mechanical Sciences, Indian Institute of Technology (IIT) Goa, Email: arindam@iitgoa.ac.in,

# INTRODUCTION.

The Slippery Liquid-Infused Porous Surface (SLIPS), inspired by the Nepenthes pitcher plant, stands out as a highly effective material for environmental and energy applications. Its unique ability to facilitate the smooth movement of various liquids enables exact control of SLIPS, which are typically created by infusing a porous substrate with a low-surface-energy lubricant[1]. This lubricant forms a stable, continuous overlayer that enables SLIPS to repel a wide variety of liquids, allowing them to glide effortlessly across the surface. Unlike traditional superhydrophobic surfaces, which are primarily effective at repelling water and often suffer from poor durability. SLIPS exhibit a broader capability to resist numerous liquids. This universal liquid-repelling characteristic enhances their functionality and addresses the limitations of superhydrophobic materials, making SLIPS a versatile and durable solution for advanced applications[1]. They have been extensively utilized in areas such as anti-biofouling, self-cleaning surfaces, liquid and mass transport systems, drag reduction, anti-corrosion coatings, and biomedical technologies[1,2,3]. Their ability to resist contamination and maintain functionality in diverse environments further enhances their appeal. Additionally, SLIPS are valued for their potential to improve energy efficiency and reduce maintenance costs in industrial and medical settings, making them a transformative solution across multiple fields.

In real-world applications, SLIPS often operate in dynamic environments where liquid droplets collide rather than remain stationary. This phenomenon of droplet impact on surfaces is common in nature and has diverse applications across various fields. Droplet impact is crucial in inkjet printing, cooling hot surfaces through sprays, operating internal combustion engines, and spraying pesticides. Droplet impact on solid surfaces involves two distinct phases: spreading and recoiling. These outcomes depend on physical parameters like kinetic energy[4], viscous forces, interfacial forces[5], etc. One such important parameter that significantly affects the outcome of droplet impact is the wettability of the surface[6,7].

Previously, numerous studies have explored the impact dynamics on SLIPS by examining the behavior of liquid droplets. These investigations focus on how droplets interact with the surface and oil, including their movement and shape changes. Baek et al.[8] analyze the impact behavior of water mixed with ethanol and glycerol at varying concentrations on SLIPS. This research highlights how surface tension and viscosity changes influence the dynamic behavior of impacting droplets on slip surfaces. In a similar study, Kim et al. [9]. investigate the impact dynamics of water droplets (diameter ~2.3 mm) falling onto slippery oil-infused surfaces, showing how wettability and surface roughness influence the droplets' behavior. The aim is to uncover. Jack Dawson et al. [10] show research which compares three surfaces: Aerogel, SLIPS, and SOCAL, focusing on differences in adhesion and friction. Aerogel, with low adhesion and friction, allows complete droplet rebound at very low Weber (We) numbers (~1). SLIPS, with high adhesion, requires higher We (>5) for a complete rebound, while SOCAL, with high adhesion and higher kinetic friction, shows no complete rebound even at We ~200. Aerogel consistently achieves 100% droplet ejection, unlike SLIPS and SOCAL. Similarly, C. Hao et al. explored the influence of surface structure, lubricant viscosity, and lubricant thickness on the impact dynamics of droplets on SLIPS (slippery liquid-infused porous surfaces). These studies provided a detailed analysis of the key factors governing droplet behaviour on SLIPS, offering valuable insights into their impact dynamics [11].

However, these studies focused only on the impact of water, since SLIPS often interact with various liquids with different properties, such as surface tension and viscosity, it is essential to study their impact behaviors. In this study, we systematically examined how surface tension and viscosity affect the impact dynamics on SLIPS by varying these properties in the liquids.

## EXPERIMENTAL SECTION.

### Fabrication of SLIPS with PDMS.

Polydimethylsiloxane (PDMS) is a widely used polymer for creating microstructures due to its adaptability and ability to be shaped into various forms. Commercially available PDMS in liquid form is mixed thoroughly with a curing agent in a 10:1 ratio (PDMS : curing agent) using a mechanical stirrer. This stirring process creates air bubbles in the mixture. To eliminate these air bubbles, the mixture is placed under a vacuum for 1 hour. After degassing, the bubble-free PDMS mixture fabricates smooth molds through a soft lithography process. In this method, the degassed PDMS is poured onto fluorosilanized smooth silicon wafers to create negative molds of the corresponding patterns (in this case, a smooth surface). The PDMS is then cured on a hot plate at 85°C for 2 hours. To prepare positive molds, the negative PDMS surfaces are functionalized with a fluorosilane coating, facilitating the easy removal of cured PDMS from both the silicon wafer and the negative PDMS mold. This process yields a positive PDMS mold that replicates the base surface, in this case, the smooth silicon wafer. The resulting smooth PDMS surface is subsequently utilized for further studies. A schematic diagram illustrating the procedure for preparing smooth PDMS for droplet impact tests is shown in Figure 1.

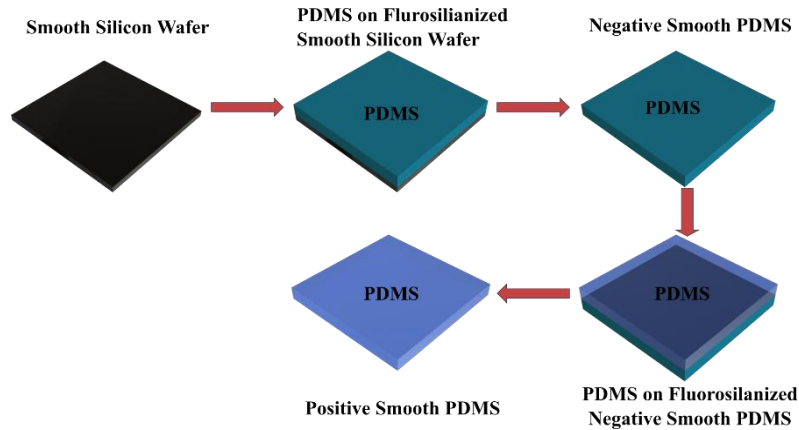

**Figure 1.** The schematic diagram outlines the procedures for preparing smooth PDMS for droplet impact tests.

As explained in many research articles[9–16], it has been observed that surface functionalization can significantly influence wettability, oil stability, and, consequently, droplet impact behavior on the surface. In this experiment, the smooth PDMS surface was also functionalized with OTS[17]. This was done to investigate the effects of wettability, oil absorption/stability, and droplet impact on the functionalized surface. The type of oil used is also a crucial parameter. So, in this experiment, we chose Silicone oil (5$cSt$), which shares the same base material as PDMS, and another Hexadecane, a hydrocarbon oil.

To investigate the effect of a thin oil layer and a bulk oil phase on droplet impact behavior, we conducted experiments on a wide range of samples with different surface functionalizations. The study began with droplet impact on smooth PDMS, followed by PDMS functionalized with OTS (PDMS OTS). To examine the effect of a thin oil layer, both PDMS and PDMS OTS surfaces were coated with silicone oil (5 $cSt$) and hexadecane, resulting in four sample sets: PDMS + SO-5$cSt$)$_{Coated}$,(PDMS + Hexadecane)$_{Coated}$, (PDMS OTS + SO-5$cSt$)$_{Coated}$,(PDMS OTS + Hexadecane)$_{Coated}$. To study the effect of a bulk oil phase, the surfaces were allowed to absorb the oils, creating another four sample sets: (PDMS + SO-5$cSt$)$_{Absorb}$,(PDMS + Hexadecane)$_{Absorb}$, (PDMS OTS + SO-5$cSt$)$_{Absorb}$,(PDMS OTS + Hexadecane)$_{Absorb}$. In total, 10 sample sets were prepared for this study to compare the effects systematically.

The oil coatings were applied using a precision dip-coating machine with a controlled withdrawal velocity of 5 mm/min[18]. For the absorption samples, the PDMS surfaces were immersed in the respective oils (silicone oil and hexadecane) and left to soak overnight. The weight of all samples was measured before and after dip-coating or absorption to assess the oil uptake. Additionally, the wettability of all samples was measured and analyzed. The droplet impact tests were conducted on all sample sets at four different Weber numbers: 28, 63, 127, and 247, representing varying impact velocities.

**Measurement of contact angles**

All samples were mounted on a goniometer (Rame Hart, Model 500) to measure Young's contact angle, advancing and receding contact angles, and droplet roll-off angles. Sessile DI water drops with a fixed volume of 8μL were vertically placed on test surfaces for contact angle measurements. Ten measurements were taken for each sample type, covering five locations per sample. Experiments were conducted at a controlled temperature of 24°C and 75% relative humidity. A monochrome video camera attached to the goniometer captured droplet images. Dynamic contact angles and contact angle hysteresis (CAH) were determined using the drop volume-changing method[19]. A needle was positioned near the surface to pump water gradually, increasing the droplet volume. The advancing contact angle was measured just before the three-phase contact line (TPCL) moved forward. Conversely, the receding contact angle was recorded during droplet suction when the TPCL began to retract. CAH was calculated as the difference between advancing and receding angles. Droplet roll-off angles were measured by placing a sessile water droplet on the surface and tilting the goniometer stage slowly until the droplet rolled off. The tilt angle at which this occurred was noted as the droplet roll-off angle.

**Droplet impact**

Droplet impact experiments were carried out by placing the sample on flat surfaces. Droplets with a diameter of 2.8 mm were produced using a Teflon-coated needle attached to a syringe, which was operated at an infusion rate of 10ml/hr using a syringe pump from Harvard Apparatus. The impact velocity of the droplets ($V_i$) was controlled by altering the fall height, which ranged from 4 cm to 35 cm, resulting in velocities between 0.88 m/s and 3.70 m/s. Preliminary tests revealed diverse droplet behaviours across this range of Weber numbers. To ensure a comprehensive analysis, five specific Weber numbers (We) were chosen: 28, 63, 127, and 245, These values span from low to high Weber numbers, representing the ratio of inertial forces to surface tension forces, defined as $We = \rho D V_i^2 / \sigma$, where $\sigma$ and $\rho$ denote the surface tension and density of water, respectively, and $D$ is the droplet diameter. The droplet impact dynamics were captured using a Phantom VEO 410 high-speed camera, which operated at a resolution of 1280×800 pixels and a frame rate of 5000 frames per second. A high-intensity light source was positioned behind the substrate, aligning the light source, substrate, and high-speed camera along the same optical axis, as illustrated in Figure 2. Video analysis was performed using MATLAB, and ImageJ software was utilized to extract detailed information from the captured images, representing different stages of droplet impact. A total 230 videos were recorded and analyzed, incorporating textured surfaces to enhance the accuracy of the droplet impact tests.

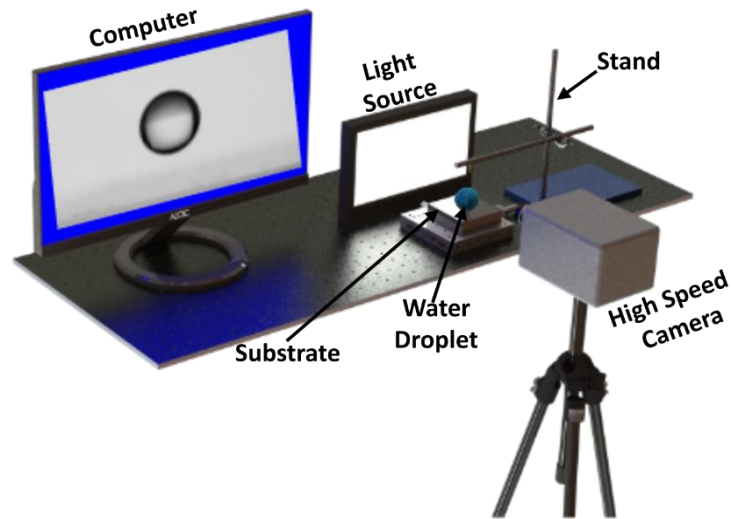

**Figure 2.** Schematic diagram of the droplet impact experimental setup.

## RESULTS AND DISCUSSIONS.

The wettability measurements for all samples were conducted with 8 repetitions to ensure repeatability. Table 1 summarizes the wettability results for the four sample types. It was observed that the contact angle hysteresis (CAH) was higher for all samples that absorbed hexadecane. Additionally, samples without OTS functionalization exhibited higher CAH compared to OTS-functionalized surfaces. These findings indicate that even after oil absorption, surface functionalization continues to play a significant role in determining wettability. The OTS-functionalized samples show a significantly lower contact angle hysteresis than the untreated samples, indicating that they are less resistant to liquid movement on their surface.

**Table 1.** Wettability of PDMS and PDMS OTS absorbed in So-5cSt and hexadecane sample, respectively.

| Functionalization | With OTS | | Without OTS | |
|---|---|---|---|---|
| Oil | SO-5$cSt$ | Hexadecane | SO-5$cSt$ | Hexadecane |
| Eq. CA | 93±3 | 91±2.5 | 94±5.5 | 87.5±3 |
| Advancing CA | 93±.5 | 91±1 | 98±1.65 | 95±1 |
| Receding CA | 90±1 | 72±1.8 | 95±1.5 | 70±2.5 |
| CAH | 2±1.2 | 15±2.1 | 3±0.9 | 20±2 |

**Table 2**. % increase in weight of PDMS and PDMS OTS absorbed in So-5cSt and hexadecane samples, respectively.

| Functionalization | With OTS | | Without OTS | |
|---|---|---|---|---|
| Oil | SO-5*cSt* | Hexadecane | SO-5*cSt* | Hexadecane |
| Increase in weight (%) | 99.96 | 43.41 | 107.15 | 66.46 |

The % increase in weight between OTS-functionalized and untreated (OTS-free) PDMS samples following impregnation with the two oils is displayed in the above table. OTS functionalization reduces weight gain for PDMS, with a more significant effect seen for Silicone Oil (99.96% increase) impregnation compared to Hexadecane (43.41% increase).

**Influence of Coating on Drop Impact Dynamics on Smooth PDMS Surface**

The droplet impact dynamics on smooth polydimethylsiloxane (PDMS) surfaces were systematically investigated across a range of Weber numbers. For all Weber numbers tested, it was consistently observed that the impacting droplets get deposited onto the surface without any rebound. This deposition behavior can be attributed to the unique combination of the surface's wettability and mechanical properties. The smooth PDMS surface exhibits moderate hydrophobicity, which, combined with its flexibility, dissipates the kinetic energy of the droplet during impact. The absence of droplet rebound is primarily due to the strong adhesion forces between the droplet and the smooth PDMS surface. Across increasing Weber numbers, the droplet's maximum spreading diameter increases and can be seen in Figure 3(a), reflecting the dominance of inertial forces at higher impact velocities. However, the eventual deposition of the droplet occurred uniformly across all the Weber number.

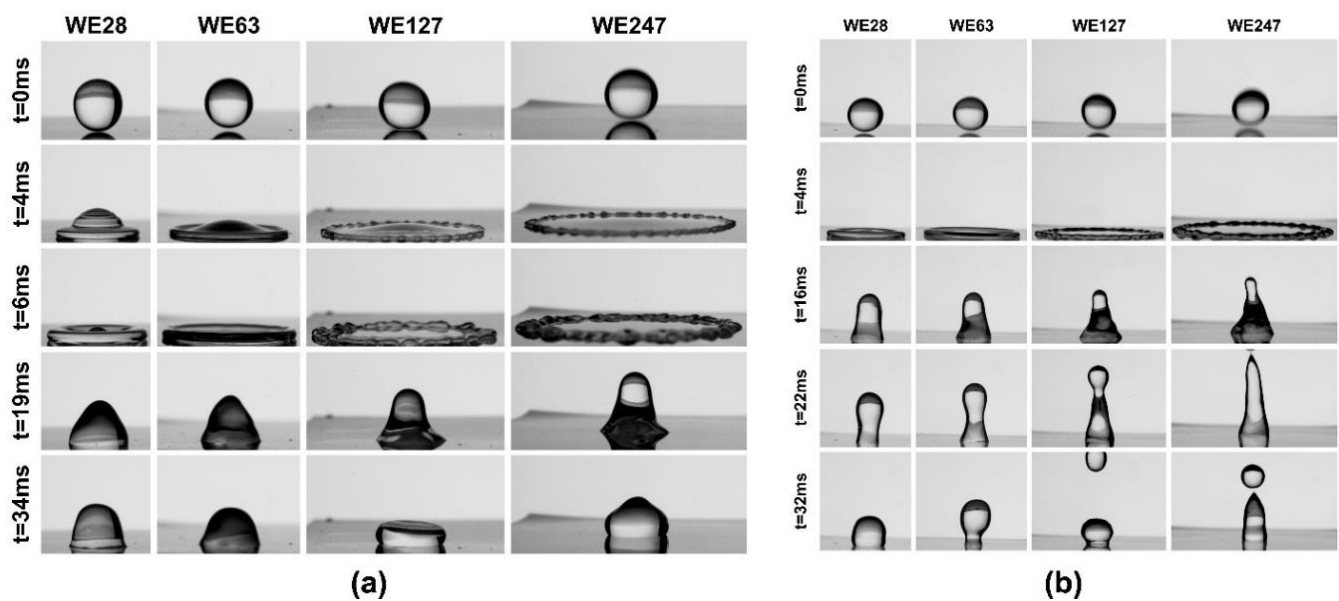

**Figure 3.** Droplet impact on (a) PDMS surface and (b) PDMS OTS surface.

The effect of functionalization on PDMS was studied by treating it with octadecyl trichlorosilane (OTS) using the method described in the experimental section. This treatment altered the surface properties, significantly changing its interaction with liquids. The modified surface exhibited reduced adhesion and increased water repellency. As observed in the contact angle measurements. When droplets were placed on the untreated PDMS surface, they spread quickly and remained in contact with it. However, after OTS treatment, the surface became significantly less sticky.

The reduction in adhesion can be attributed to the presence of the OTS layer, which introduces a low-energy, hydrophobic surface. This modification decreases the contact angle hysteresis and weakens the pinning forces at the three-phase contact line, enhancing the droplet's ability to partially rebound post-impact. Furthermore, the maximum spreading diameter of the droplet during impact was observed to be increased on the OTS-functionalized surface, suggesting a decrease in the energy dissipation caused by surface adhesion. Another notable effect of OTS functionalization was the formation of taller droplet pillars during the recoiling phase (see Figure 3(b). Compared to the unmodified smooth PDMS surface, the increased droplet height reflects the improved elasticity and capillary interactions of the recoiling droplet with the low-adhesion surface. These effects were more pronounced at higher Weber numbers, where the inertial forces dominate the impact dynamics. The results demonstrate that OTS functionalization effectively transforms the wetting behaviour of smooth PDMS, enabling partial rebound and reduced droplet deposition. This transition is crucial for applications requiring reduced contact time and minimal liquid retention, such as self-cleaning surfaces and liquid transport systems.

As shown in Figure 4, which illustrates the droplet impact of water at different Weber numbers on a temporal scale, the samples used are PDMS surfaces functionalized with OTS and subsequently coated with oil, as described in the experimental section. The results indicate that for all Weber numbers, the maximum spreading diameter of the droplet occurs on the PDMS-OTS surface, compared to surfaces coated with SO-5*cSt* and hexadecane. The reduced droplet spreading on oil-coated samples can be attributed to the viscous nature of the oil, which resists the flow of the droplet. Moreover, the oil does not form a distinct surface layer but is absorbed into the substrate, leading to localized changes in surface flexibility. This altered flexibility further restricts the droplet's ability to spread. These combined effects—viscous resistance and changes in surface dynamics reason for the observed differences in droplet spreading behavior across the samples.

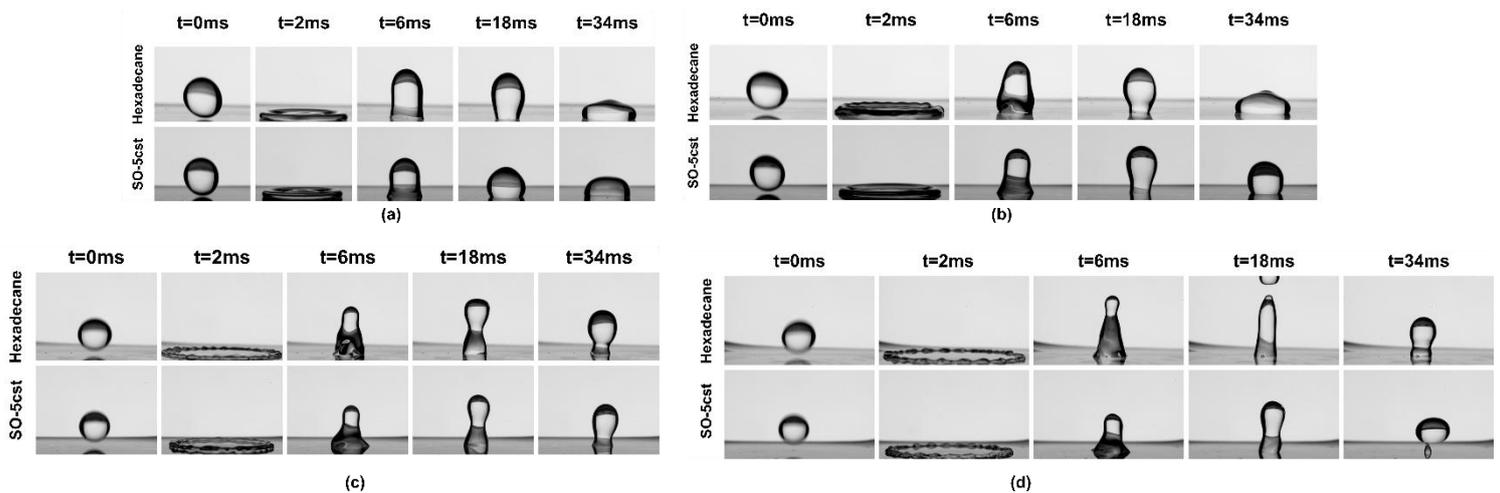

**Figure 4.** Droplet impact on PDMS OTS samples when coated with a different oil for the different Weber numbers (a) *We*=28 (b) *We* =63(c) *We*=127 (d) *We* =245.

When PDMS is not functionalized with OTS and then subsequently coated with a layer of oil, the oil coating does not significantly influence the droplet's rebound behavior or spreading diameter compared to the non-coated PDMS surface. This suggests that the presence of the oil layer, while modifying the surface properties, does not impose sufficient resistance to alter the droplet's dynamics upon impact. Shown in Figure 5. The spreading and rebound characteristics remain governed primarily by the inherent properties of the underlying PDMS substrate. The absorbed nature of the oil layer likely prevents it from acting as a distinct barrier to the droplet's motion, resulting in behavior similar to that of an uncoated PDMS surface. Thus, the oil coating appears to have a minimal role in modifying the impact dynamics

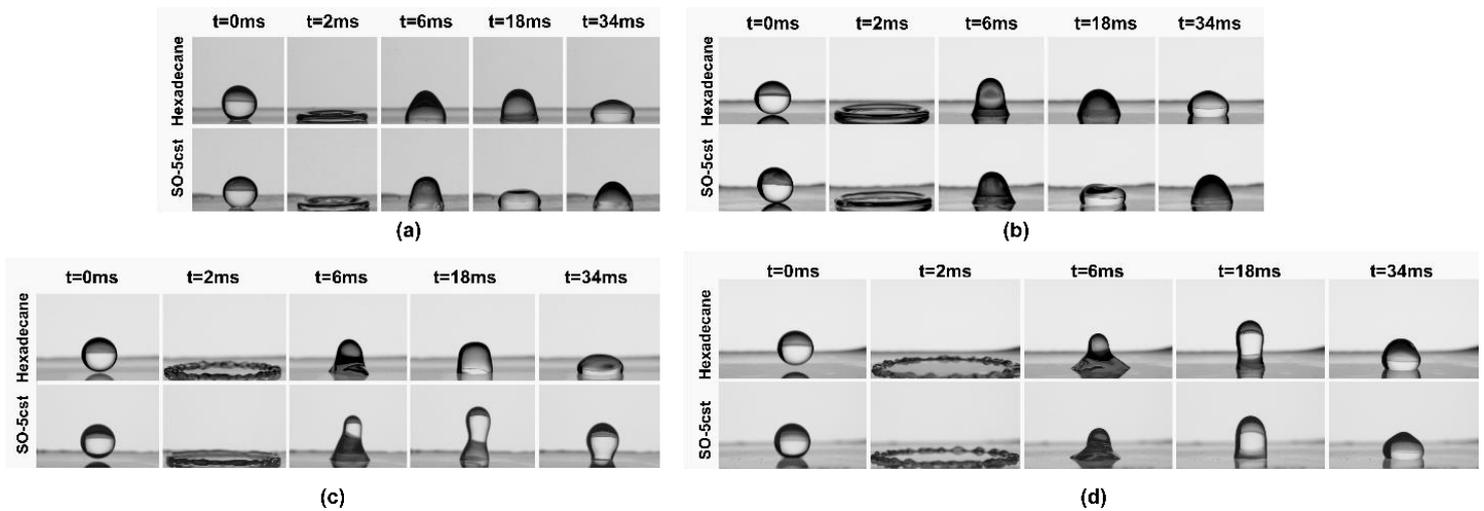

**Figure 5.** Droplet impact on PDMS samples when coated with a different oil for the different Weber numbers (a) *We*=28 (b) *We* =63 (c) *We*=127 (d) *We* =245.

**Influence of Oil Absorption on Drop Impact Dynamics on Smooth PDMS Surface**

To investigate the absorption of oil into PDMS, we immersed both PDMS and PDMS-OTS samples in SO-5*cSt* and hexadecane overnight. The changes in size and weight of the samples were measured before and after absorption. It was observed that SO-5*cSt* was absorbed more extensively in both PDMS and PDMS-OTS samples, as evidenced by a higher percentage increase in weight compared to hexadecane. This indicates that SO-5*cSt* exhibits greater compatibility with the PDMS matrix. Following the absorption experiment, droplet impact tests were conducted on all four sample variations—PDMS and PDMS-OTS samples absorbed in SO-5*cSt* and hexadecane—at different Weber numbers, as shown in Figures 6 and 7. For the samples absorbed with SO-5*cSt*, both PDMS and PDMS-OTS consistently exhibited full droplet rebound at all Weber numbers. This rebound behavior highlights the impact of the absorbed oil on the surface dynamics.

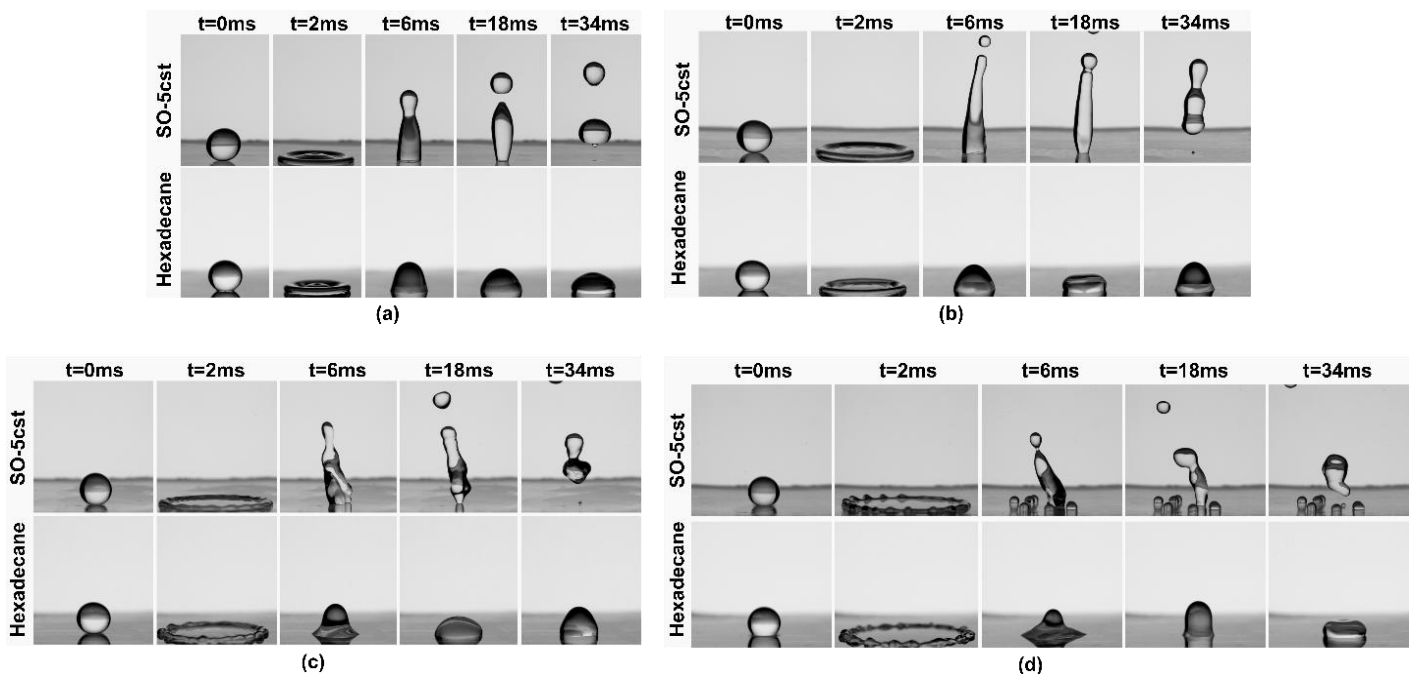

**Figure 6.** Droplet impact on PDMS samples when absorbed with a different oil for the different Weber numbers (a) *We*=28 (b) *We* =63 (c) *We*=127 (d) *We* =245.

In contrast, the PDMS sample absorbed in hexadecane did not exhibit droplet rebound, and the spreading diameter was smaller compared to PDMS absorbed in SO-5*cSt*. This difference can be attributed to weaker molecular interactions between PDMS and hexadecane, which prevent the formation of a thin oil film on the outermost layer of the PDMS surface. Although hexadecane is absorbed into the PDMS matrix, it does not remain localized on the top layer, reducing its ability to influence droplet dynamics effectively.

Another possible explanation is the alteration in surface flexibility caused by hexadecane absorption. This change in flexibility appears to resist droplet motion, further limiting spreading and preventing rebound. On the other hand, the PDMS-OTS sample absorbed in hexadecane exhibited behavior similar to the unabsorbed PDMS-OTS sample. This observation confirms that hexadecane absorption in both PDMS and PDMS-OTS does not create a uniform thin oil layer on the surface, which is essential for modifying impact dynamics. Supporting evidence from wettability results aligns with these findings. The CAH values for hexadecane-absorbed PDMS and PDMS-OTS samples were found to be high, indicating a lack of significant surface lubrication. These results collectively demonstrate that the behavior of hexadecane-absorbed samples is distinct from SO-5*cSt*-absorbed samples, highlighting the critical role of oil type and its interaction with the PDMS matrix in influencing droplet impact and spreading phenomena.

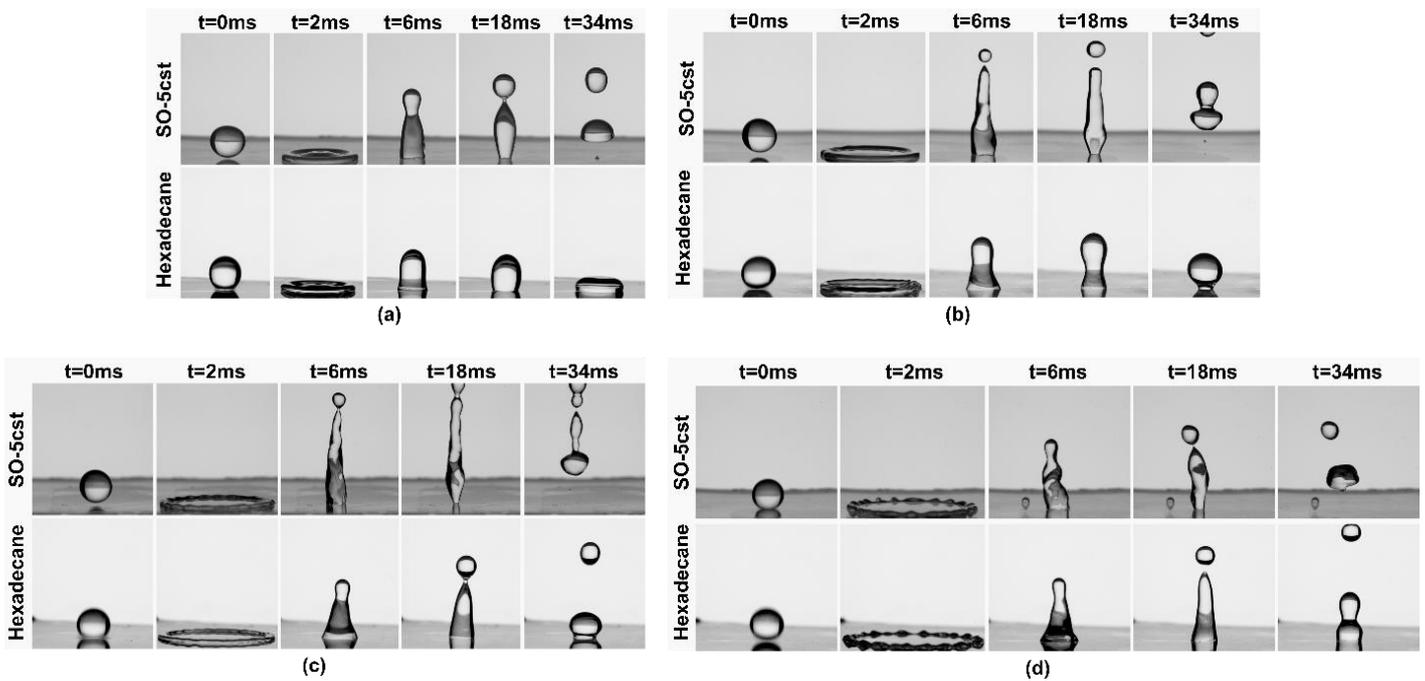

**Figure 7.** Droplet impact on PDMS OTS samples when absorbed with a different oil for the different Weber numbers (a) $We$=28 (b) $We$=63 (c) $We$=127 (d) $We$=245.

## CONCLUSION

The study highlights the critical influence of surface functionalization, oil type, and application methods on the wettability and droplet impact behavior of PDMS surfaces. The results demonstrate that OTS-functionalized PDMS (PDMS+OTS) surfaces exhibit significantly lower CAH than untreated PDMS, resulting in enhanced hydrophobicity and reduced adhesion. This functionalization allows for partial droplet rebound during impact, particularly at higher Weber numbers, indicating its effectiveness in modifying surface interaction dynamics. Oil-coated surfaces further revealed interesting dynamics. The viscous nature of the oils, particularly silicone oil (SO-5*cSt*) and hexadecane, reduced droplet spreading. The coating behavior underscores the importance of oil viscosity and its ability to form a distinct surface layer in influencing droplet spreading and adhesion.

The absorption experiments highlighted the differences between SO-5*cSt* and hexadecane in their interaction with PDMS. SO-5*cSt* was absorbed more extensively, forming a lubricating layer that significantly reduced adhesion and allowed for consistent droplet rebound. On the other hand, hexadecane's weaker compatibility with PDMS resulted in stiffer surfaces, limited droplet spreading, and the absence of rebound. These differences highlight the critical role of oil type in defining the surface behavior of PDMS. Interestingly, oil absorption also influenced surface flexibility, which in turn affected droplet dynamics. SO-5*cSt* absorption maintained a lubricated surface, enabling full droplet rebound, while hexadecane absorption increased surface stiffness and reduced droplet mobility. Even after oil absorption, OTS-functionalized surfaces exhibited better performance, showing that surface functionalization continues to play a vital role in modifying wettability and droplet impact behavior. Overall, the study concludes that combining OTS functionalization and SO-5*cSt* absorption is the most effective approach for minimizing droplet adhesion and enhancing rebound.

## ACKNOWLEDGMENTS

The authors want to thank the School of Mechanical Sciences and Centre of Excellence in Particulates, Colloids and Interfaces, Indian Institute of Technology Goa, for providing the experimental facility and necessary support to conduct the above work. We would also like to thank Yash Khobragade. This research work was funded by the Science and Engineering Research Board with Sanction Order Nos. CRG/2023/008620.

## References

[1] J. Li, B. Lu, Z. Cheng, H. Cao, and X. An, "Designs and recent progress of 'pitcher plant effect' inspired ultra-slippery surfaces: A review," Prog. Org. Coatings **191**(April), 108460 (2024).

[2] J.L. Yin, M.L. Mei, Q.L. Li, R. Xia, Z.H. Zhang, and C.H. Chu, "Self-cleaning and antibiofouling enamel surface by slippery liquid-infused technique," Sci. Rep. **6**(April), 1–14 (2016).

[3] Q. Ma, W. Wang, and G. Dong, "Facile fabrication of biomimetic liquid-infused slippery surface on carbon steel and its self-cleaning, anti-corrosion, anti-frosting and tribological properties," Colloids Surfaces A Physicochem. Eng. Asp. **577**(May), 17–26 (2019).

[4] J. Eggers, M.A. Fontelos, C. Josserand, and S. Zaleski, "Drop dynamics after impact on a solid wall: Theory and simulations," Phys. Fluids **22**(6), 1–14 (2010).

[5] H. Wang, H. Lu, and W. Zhao, "A review of droplet bouncing behaviors on superhydrophobic surfaces: Theory, methods, and applications," Phys. Fluids **35**(2), (2023).

[6] M. Broom, and G.R. Willmott, "Water drop impacts on regular micropillar arrays: The impact region," Phys. Fluids **34**(1), (2022).

[7] S.H. Kim, H. Seon Ahn, J. Kim, M. Kaviany, and M. Hwan Kim, "Dynamics of water droplet on a heated nanotubes surface," Appl. Phys. Lett. **102**(23), (2013).

[8] S. Baek, and K. Yong, "Impact dynamics on SLIPS: Effects of liquid droplet's surface tension and viscosity," Appl. Surf. Sci. **506**(November 2019), (2020).

[9] S. Kim, T. Wang, L. Zhang, and Y. Jiang, "Droplet impacting dynamics on wettable, rough and slippery oil-infuse surfaces," J. Mech. Sci. Technol. **34**(1), 219–228 (2020).

[10] J. Dawson, S. Coaster, R. Han, J. Gausden, H. Liu, G. McHale, and J. Chen, "Dynamics of Droplets Impacting on Aerogel, Liquid Infused, and Liquid-Like Solid Surfaces," ACS Appl. Mater. Interfaces **15**(1), 2301–2312 (2023).

[11] M. Kanungo, S. Mettu, K.Y. Law, and S. Daniel, "Effect of roughness geometry on wetting and dewetting of rough PDMS surfaces," Langmuir **30**(25), 7358–7368 (2014).

[12] A.S. Vaillard, M. Saget, F. Braud, M. Lippert, L. Keirsbulck, M. Jimenez, Y. Coffinier, and V. Thomy, "Highly stable fluorine-free slippery liquid infused surfaces," Surfaces and Interfaces **42**(PA), 103296 (2023).

[13] B. He, W. Chen, and Q. Jane Wang, "Surface texture effect on friction of a microtextured poly(dimethylsiloxane) (PDMS)," Tribol. Lett. **31**(3), 187–197 (2008).

[14] A. Alizadeh, V. Bahadur, W. Shang, Y. Zhu, D. Buckley, A. Dhinojwala, and M. Sohal, "Influence of substrate

elasticity on droplet impact dynamics," Langmuir **29**(14), 4520–4524 (2013).

[15] L. Chen, E. Bonaccurso, P. Deng, and H. Zhang, "Droplet impact on soft viscoelastic surfaces," Phys. Rev. E **94**(6), 1–9 (2016).

[16] L.Z. Wang, A. Zhou, J.Z. Zhou, L. Chen, and Y.S. Yu, "Droplet impact on pillar-arrayed non-wetting surfaces," Soft Matter **17**(24), 5932–5940 (2021).

[17] S.S. Ganar, and A. Das, "Experimental insights into droplet behavior on Van der Waals and non-Van der Waals liquid-impregnated surfaces," (December 2022), (2024).

[18] S. Li, F. Zhao, Y. Bai, Z. Ye, Z. Feng, X. Liu, S. Gao, X. Pang, M. Sun, J. Zhang, A. Dong, W. Wang, and P. Huang, "Slippery liquid-infused microphase separation surface enables highly robust anti-fouling, anti-corrosion, anti-icing and anti-scaling coating on diverse substrates," Chem. Eng. J. **431**(P1), 133945 (2022).

[19] S.S. Ganar, and A. Das, "Unraveling the interplay of leaf structure and wettability: A comparative study on superhydrophobic leaves of Cassia tora, Adiantum capillus-veneris, and Bauhinia variegata," (February), (2024).